\documentstyle[12pt]{article}
\begin{document}
\newcommand{\ad}[1]{\mbox{${#1}^{\dagger}$}}
\newcommand{\kone}[1]{\mbox{$|{#1}\rangle$}}
\newcommand{\ktwo}[2]{\mbox{$|{#1}_{#2}\rangle$}}
\newcommand{\kthree}[3]{\mbox{$|{#1}\,;\,{#2}_{#3}\rangle$}}
\newfont{\blackb}{msbm10 scaled 1200}
\def\BR{\mbox{\blackb R}}
\def\BC{\mbox{\blackb C}}

\title{Ladder operators for isospectral oscillators}
\author{S. Seshadri, V. Balakrishnan and S. Lakshmibala\\
{\em Department of Physics, Indian Institute of Technology,}\\
{\em Madras 600 036,
India}}
\date{}
\maketitle
\begin{abstract}

We present, for the isospectral family of oscillator Hamiltonians, 
a systematic procedure for constructing raising and
lowering operators satisfying any prescribed `distorted' Heisenberg
algebra (including the $q$-generalization). This is done by means
of an operator transformation implemented by a  shift operator. The latter
is obtained by solving an appropriate partial isometry condition in
the Hilbert space.  Formal  representations of the non-local
operators concerned are given in terms of pseudo-differential operators.
Using the new annihilation operators, new classes of 
coherent states are constructed for isospectral oscillator
Hamiltonians. The corresponding Fock-Bargmann representations are also
considered, with specific reference to the order of the entire function
family in each case.
\end{abstract}
\hskip 1cm PACS Nos.\,\, 03.65.Fd, 02.30.Tb, 03.65.Bz, 42.50.-p 
\newpage
\section*{\large I. INTRODUCTION}
\renewcommand{\theequation}{1.\arabic{equation}}
\setcounter{equation}{0} 
\hskip .5cm Generalized coherent states of different kinds 
have been defined and investigated for some years now,
particularly in quantum optics and quantum measurement theory.$\,^{
1-6}$\\ Recently, two different kinds of coherent states (CS) have
been constructed$\,^{ 7,8}$ in relation to a one-parameter  $(\lambda
\in \BR)$ family of Hamiltonians $\widetilde H$\,  that are
isospectral companions of the standard oscillator Hamiltonian $H=
a^{\dagger}a$. (The latter corresponds to the limit $|\lambda|
\rightarrow\infty $.) For each finite value of $\lambda$, the eigenstates
of $\widetilde H$ \,  consist of a zero-energy state \ktwo{\theta}{0} and
a set of states \{\ktwo{\theta}{n}\} with eigenvalues
$n=1,2,\ldots$, the {\em latter} being obtained from the oscillator states
\kone{n} ($n= 0,1,\ldots$) by a non-unitary transformation (see
Eq. (2.5) below). Fernandez {\em et al}.$\,^7$ have constructed coherent
states that are eigenstates of an annihilation operator associated
with  \{\ktwo{\theta}{n}, $ n\geq 1$\}.  However, these states are
not coherent states in the group-theoretic or Perelomov sense,$\,^9$
since they are not obtained by the action of a unitary `displacement'
operator on the base state \ktwo{\theta}{1}. On the other hand, Kumar
and Khare$\,^8$ have shown that $\widetilde H$\,  itself is essentially
unitarily equivalent to $H$. This fact enables them to obtain
coherent states simply by a unitary transformation of the standard
oscillator CS. These states are thus coherent states in the
group-theoretic sense in addition to being annihilation operator
eigenstates. As may be anticipated, they reduce to the standard
oscillator CS in the limit $|\lambda|\rightarrow\infty$.
Subsequently, Fernandez {\em et al.}$\,^{10}$ have generalized their earlier
work$\,^7$ by introducing a real parameter $w$ as follows: 
the commutator of the lowering and raising operators in the
\{\ktwo{\theta}{n}, $n \ge 0$\} basis is a diagonal operator given by
 diag (0,$w$,1,1,1,...). This has been termed a `distorted' Heisenberg 
 algebra. In the special case $w$ = 1, the CS constructed are also
 CS in the group-theoretic sense. 

A number of interesting problems  arise, which we address in this
paper: the generalization of the algebra to encompass an arbitrary
diagonal operator as the basic commutator, including the
corresponding $q$-generalization; the development of a
systematic procedure to find raising and lowering operators in this
case; and the construction of the corresponding CS (which
are {\em not} unitarily equivalent to the standard
oscillator CS). Our strategy is to find a suitable operator transformation
that takes the original ladder operators to the required new ones.
The transformation is implemented by means of a {\em shift operator}
that is constructed so as to satisfy  appropriate partial isometry
conditions dictated by the algebra imposed. We also derive
representations for the new raising and lowering operators in terms
of pseudo-differential operators, to enable one to compare their
structure explicitly with their  original counterparts. Finally, we
examine the Fock-Bargmann (or entire analytic function)
representation of states based on the CS obtained in various cases,
with particular reference to the order of the entire function
families concerned.

\section*{\large II. NOTATION AND REVIEW} 
\renewcommand{\theequation}{2.\arabic{equation}}
\setcounter{equation}{0}
\hskip .5cm For ready reference, we recapitulate briefly the relevant parts of
Refs. 7, 8 and 10. Given $H=a^{\dagger}a$ with [$a,a^{\dagger}]=1$, we have
$H\kone{n}=n$\kone{n} ($n=0,1,\ldots$). In the
position representation, $a$ = $2^{^{-1/2}}\,(x+d\,)$ and  $a^{\dagger}$ = 
$2^{^{-1/2}}\,(x-d\,)$, where 
$d\,=\, d/dx$. Operators $b$ and $b^{\dagger}$ are then defined,
 given in the position representation by 
\begin{equation}
b\,=\, 2^{-\frac{1}{2}}\,( x + d + \phi (x))\:\;,\:\; 
b^{\dagger}\, =\,2^{-\frac{1}{2}} ( x - d + \phi(x))\:,
\end{equation} and the condition $bb^{\dagger}=aa^{\dagger}$ imposed. This
implies that the function $\phi(x)$ satisfies the Riccati equation
\begin{equation}
\phi\/' + 2x\phi + \phi^{^2} = 0 
\end{equation}
where $\phi\/' = d\phi/dx$. Equation\,(2.2) has the one-parameter
family of solutions
\begin{equation}
\phi_{\lambda} (x)=\frac{e^{-\,x^2}}{ \left
({\lambda\,+\int^x_0\;dy\;e{^{-\,y^2}}} \right )}
\end{equation}
where $|\lambda|\; > \;{\sqrt{\pi}}/{2}$ (to avoid a
singular $\phi_\lambda\,$). Now consider the Hamiltonian (actually, a
family of Hamiltonians, parametrized by $\lambda\; \in
\;(-\infty\,,\,\infty)$) 
\begin{equation}
\widetilde H=b^{\dagger}b\,.
\end{equation}
In order to keep the notation simple, the $\lambda$-dependence of
$b\,,\,b^{\dagger}$ and $\widetilde H$ has not been
indicated explicitly. Although $bb^{\dagger}=aa^{\dagger}$  (that is, the
supersymmetric partners of $H$ and $\widetilde H$\,  are identical to each other),
$b^{\dagger}b\,\neq\,a^{\dagger}a$. It is straightforward to check that the 
states 
\begin{equation}
\ktwo{\theta}{n}=n^{-{\frac{1}{2}}}\,\,b^{\dagger}\,\kone{n-1}\:\:(n=1,2,\ldots)
\end{equation}
are eigenstates of the Hamiltonian $\widetilde H$ \, with 
eigenvalues\, $n=1,2,\ldots$. The {\em ground state}
\ktwo{\theta}{0}  of $\widetilde H$ \, is obtained by requiring
that 
\begin{equation}
b\,\ktwo{\theta}{0}=0\:.
\end{equation}
The eigenstates \{\ktwo{\theta}{n}, $n=0,1,\ldots$\} form a
complete orthonormal set of states in the Hilbert space $\cal H$
spanned by the Fock basis \{\kone{n}\}. It will be useful
to write ${\cal H}_0 = {\mbox{span}}\,\kone{0}\,,\,{\cal H}_1
=\,{\mbox{span}}\,\{\kone{n} , n \geq 1\}\,,\;{\widetilde {\cal H}}_0 =\,
{\mbox{span}}\,\ktwo{\theta}{0}$ and ${\widetilde {\cal H}}_1
=\,{\mbox{span}}\,\{\ktwo{\theta}{n} , n \geq 1\}$, so that ${\cal H} =
{\cal H}_0 + {\cal H}_1 = {\widetilde {\cal H}}_0 + {\widetilde {\cal H}}_1 $.
(For easy identification, we shall use a tilde for spaces and
operators pertaining to the isospectral case and drop
the tilde for their counterparts relating to the standard oscillator.)
  It is evident that $H$ and $\widetilde H$ \,
are isospectral, and that $\widetilde H$ $\,\rightarrow H$ in the limit
$|\lambda|\rightarrow \infty$. However, since $b^{\dagger}b$ $\neq
a^{\dagger}a$, the commutator [$b,\ad{b}$] $\neq 1$. Thus \ad{b} and
$b$ are {\em not} the appropriate raising and lowering operators
for ${\widetilde H}$ . Fernandez {\em et al}.$\,^7$ consider
the lowering operator (in their notation)
\begin{equation}
A=\ad{b}a\,b
\end{equation} 
which annihilates not only \ktwo{\theta}{0} but also \ktwo{\theta}{1} ; one
finds
\begin{equation}
A\ktwo{\theta}{0}=0\:\:,\:\:
A\ktwo{\theta}{n}=(n-1)\,n^{\frac{1}{2}}\,\ktwo{\theta}{n-1}.
\end{equation}
Similarly,
\begin{equation}
\ad{A}\,\ktwo{\theta}{0}=0\:\:,\:\:\ad{A} \ktwo{\theta}{n}
=n(n+1)^{\frac{1}{2}}\ktwo{\theta}{n+1}.
\end{equation}
These relations help us understand how the Hilbert space $\cal
H$ breaks up in a natural fashion into the direct sum
${\widetilde {\cal H}}_0 + {\widetilde {\cal H}}_1$.  As $A$ is an
annihilation operator in ${\widetilde {\cal H}}_1$, its eigenstates,
satisfying
$A\,\kthree{z}{\theta}{1}\,=\,z\,\kthree{z}{\theta}{1}\,\,(z\,\in\,\BC
)$, may be regarded as coherent states. They are given by$\,^7$
\begin{equation}
\kthree{z}{\theta}{1}=\:{\mbox{(const.)}}\:\sum_{n=0}^{\infty}\:\frac{z^{n}}{
n! \sqrt{(n+1)!}}\,\ktwo{\theta}{n+1}
\end{equation} 
However, since
[$A\,,\,\ad{A}]\,\neq\,1$ even in the subspace ${\widetilde {\cal
H}}_1$, the state \,\kthree{z}{\theta}{1} cannot be obtained by
a unitary transformation of the form exp$\,(z \ad{A} - {\overline
z}A$) acting on the ground state
\ktwo{\theta}{1}.  Therefore \kthree{z}{\theta}{1} is not a CS
in the group-theoretic sense, nor can {\em generalized} coherent
states be built up from it by the standard procedure -- namely,
by using \kthree{z}{\theta}{1} as the ground state of an
appropriate Hamiltonian obtained by a unitary transformation of
$\widetilde H$.

In contrast to the foregoing, Kumar and Khare$\,^8$ have shown that $b$ and
\ad{b} can be written in the form
\begin{equation}
b=a\ad{U}\,\,,\,\,\ad{b}=U\ad{a}
\end{equation}
where $U$ is unitary ($U\ad{U}=\ad{U} U = 1$) in the full
space ${\cal H}$\,, and formally determined the matrix elements of $U$
in the basis
\{\kone{n}\}. Therefore $b\ad{b}\,=\,a\ad{a}$, as required.
Further, $\widetilde H$ $=\ad{b}b=U H \ad{U}$, so that $\widetilde H$ \,and $H$ are actually
unitarily equivalent. Writing ${\tilde a}=U a\, \ad{U}$, one has $\widetilde H$ 
$\,=\,\ad{\tilde a} {\tilde a}$ where [${\tilde a}\,,\,\ad{\tilde a}$] = 1. The
eigenstates \ktwo{\theta}{n} of $\widetilde H$ \, are just the unitary transforms of
those of $H$,\, $i.e.$,
\begin{equation}
 \ktwo{\theta}{n} =  U \,\kone{n}\;\; {\mbox{and}}\;\;
\ad{U}\,\ktwo{\theta}{n} =  \kone{n},\; n=0,1,\ldots \
\end{equation} .
The operator $\tilde a$ only annihilates the vacuum
\ktwo{\theta}{0}. The state
\begin{equation}
\kthree{\alpha}{\theta}{0}= \exp \left(-\frac{\,|\alpha|^2}{2}  \right) \;
\sum_{n=0}^{\infty}\,\,\frac{\alpha^{n}} {\sqrt{n!}}\,\,\ktwo{\theta}{n}\;\;\;
(\alpha \in {\BC})
\end{equation} 
is both an annihilation operator eigenstate and a CS in the
Perelomov sense.  It is readily verified that (i)
\kthree{\alpha}{\theta}{0} = $U\,\,\kone{\alpha}$ where
\kone{\alpha} is the standard oscillator CS; (ii)
\kthree{\alpha}{\theta}{0} is obtainable by a unitary displacement operator
acting on \ktwo{\theta}{0}; and  (iii) \kthree{\alpha}{\theta}{0}
$\rightarrow$
\kone{\alpha} in the limit $|\lambda|\,\rightarrow \,\infty$. Generalized CS
may be constructed from \kthree{\alpha}{\theta}{0} in essentially the same
manner as in the case of \kone{\alpha}. Viewed in this manner, the isospectral 
case is simply a unitary transformation of the standard oscillator problem,
and nothing new is gained by considering it in preference to the latter.

In order to have ladder operators in  ${\widetilde {\cal H}}_1$ 
 whose commutator is equal to unity, and also to generalize this
algebra to some extent, Fernandez {\em et al}.\ 
have  subsequently defined$\,^{10}$ a new lowering operator $C_w$ (in
their notation) according to 
\begin{equation}
C_w=\ad{b}f(H)\,a\,b.
\end{equation}
Here, the form of $f(H)$ is deduced from the requirement that the
commutator $[C_w,C_w^\dagger]$ be equal to diag ($w,1,1,...$) in  ${\widetilde {\cal H}}_1$, where $w$ is any positive number. (The operators $C_w$ and
$C_w^\dagger$ continue to annihilate the subspace ${\widetilde {\cal H}}_0$.)
 It is
found that
\begin{equation}
f(H) = \frac{1}{H+1} \,\left( \frac{H+w}{H+2} \right)^{\frac{1}{2}} .
\end{equation}
It is evident that diag ($w,1,1,...$) is not the most general diagonal
form possible for the basic commutator, although it can be shown to be
the only consistent one once the specific form given by Eq. (2.14) is
assumed for the lowering operator. As we shall see, a procedure of
wider applicability is to look for an operator
 transformation on $\tilde a$  of the type $\ad{{\tilde S}}{\tilde a}{\tilde S}$. 
 In the next section,
we develop a systematic procedure to find the transformation $\tilde
S$, and from it the ladder operators {\em for any prescribed diagonal form
of their commutator}, including the case of the 
$q$-generalization of the isospectral oscillator. 

\section*{\large III. SHIFT OPERATORS AND OPERATOR TRANSFORMATIONS}
\renewcommand{\theequation}{3.\arabic{equation}}
\setcounter{equation}{0}

\hskip .5cm We wish to implement transformations on the ladder operators ${\tilde
a}$ and 
$\ad{\tilde a}$ so as to obtain a lowering operator
\begin{equation}
\tilde a_1 = \ad{{\tilde S}}\,{\tilde a}\,{\tilde S}
\end{equation}
and a hermitian conjugate raising operator 
\begin{equation}
\ad{\tilde a_1} = \ad{{\tilde S}}\,{\ad{\tilde a}}\,{\tilde S}
\end{equation}
with the following properties:
both operators must annihilate $\ktwo{\theta}{0}$
(so that ${\widetilde {\cal H}}_0$ is the kernel of the
transformation); and 
their commutator is a prescribed diagonal operator in 
${\widetilde {\cal H}}_1$. For the sake of clarity, we solve 
the problem first in
the standard Fock basis, and then use the operator $U$ to
 transform to the isospectral case.
 
Accordingly, we first seek to eliminate the subspace ${\cal H}_0$ from ${\cal
H}$, and to find conjugate lowering and raising operators in  ${\cal
H}_1$ given by
\begin{equation}
 a_1 = \ad{S}\, a\, S\;\;,\;  a_1^\dagger = \ad{S}\, \ad{a}\,S,
 \end{equation}
   whose commutator is equal to  {\em unity} throughout
the latter space. For this we need to find an operator $S$  in ${\cal H}$ that
 satisfies the {\em
partial isometry}
\begin{equation}
S\ad{S} = 1 ,\; 
\ad{S}S = 1 - \kone{0} \langle 0 |.
\end{equation}
 It is easy to see that this is solved by the {\em shift} operator
 \begin{equation}
S =
\sum_{n=0}^{\infty} \; \kone{n} \langle n+1 |.
\end{equation}
(Note that this operator has the closed form
 $S = (1+ H)^{-1/2}\,a$ in ${\cal H}$.) It is easily verified that $a_1$ and
$a_1^\dagger$ satisfy
\begin{equation}
a_1\,\kone{0}\,=\,a_1^\dagger\,\kone{0}\,=\,0\;,\;a_1\,\kone{1}\,=\,0.
\end{equation} 
 Now consider the {\em general} `distorted' Heisenberg algebra 
 in which the commutator is specified (in the Fock basis) to be
\begin{equation}
[a_1,a_1^\dagger] =\; \mbox{diag}\,(0,w_1,w_2,...) =\; \sum_{n=1}^{\infty} \; w_n\, 
\kone{n} \langle n |,
\end{equation}
where $\{w_n\}$ are real positive constants. $S$ is now found by
generalising the partial isometry condition, Eq. (3.4), to read 

\begin{equation}
S\ad{S} = \sum_{n=0}^{\infty} \; c_n \,\kone{n} \langle n |\;,\;\;  
\ad{S}S = \sum_{n=0}^{\infty} \; c_n \,\kone{n+1} \langle n+1 |,  
\end{equation}
where the coefficients $c_n$ are to be determined. The deviation of
the $c_n$'s from unity is a measure of the extent to which the shift
operator deviates from the exact partial isometry specified in Eqs. (3.4).
Solving for $S$
and $\ad{S}$ in the form
\begin{equation}
S = \sum_{n=0}^{\infty} \; c_n^{^{\frac{1}{2}}} \,\kone{n} \langle n+1 | \;,\; 
\ad{S} = \sum_{n=0}^{\infty} \; c_n^{^{\frac{1}{2}}}\, \kone{n+1} \langle n |  
\end{equation}
and using Eqs. (3.3) and (3.7), we obtain the recursion relations 
\begin{equation}
c_0c_1\,=\,w_1\;\;,\;\;(n+1)\,c_nc_{n+1}\,-n\,c_nc_{n-1}\,=\,w_{n+1}\;\;(n=1,2,3,...).
\end{equation}
We may choose $c_0 = 1$ without loss of generality, and solve the
recursion relations to get 
\begin{equation}
c_n = \frac{(n-1)!!}{n!!} 
\frac{(w_1+w_2+\cdots+w_n)!!}{(w_1+w_2+\cdots+w_{n-1})!!}.
\end{equation}
Here the notation $(w_1+w_2+\cdots+w_n)!!$ stands for the product
\begin{eqnarray}
(w_1+w_2+\cdots+w_n)\times(w_1+w_2+\cdots+w_{n-2})\cdots
\nonumber
\end{eqnarray}
Substituting the values for $c_n$ given above in Eq. (3.9) yields
 $S$ and $\ad{S}$. The corresponding expressions for $a_1$ and $a_1^\dagger$
are then obtained from Eqs. (3.3), and their commutator is guaranteed
to satisfy Eq. (3.7).

We go over now to the isospectral case. As the corresponding
shift and ladder operators are merely the unitary transformations
\begin{equation}
\tilde S \,=\, U\, S\, \ad{U}\;,\;\, 
{\tilde a}_1 \,=\, U\, a_1\, \ad{U}\;,\;\,
{\ad{\tilde a_1}} \,=\, U\, a_1^\dagger \,\ad{U}\;,
\end{equation}
we find  
\begin{eqnarray}
\widetilde S &=&  \,\sum_{n=0}^{\infty} \, c_n^{^{\frac{1}{2}}} \,\, \ktwo{\theta}{n}
\langle {\theta}_{n+1} |, 
\end{eqnarray}
where $c_0 = 1$ as already stated, and $c_n$ is given by Eq. (3.11); and
\begin{eqnarray}
{\tilde a}_1 \;&=&\;\sum_{n=1}^{\infty}\;{(w_1+w_2+\cdots+w_n)}^{\frac{1}{2}}\,
\,\ktwo{\theta}{n} \langle \theta_{n+1}|\;,\\
\;\ad{{\tilde a_1}} \;&=&\;\sum_{n=1}^{\infty}\;{(w_1+w_2+\cdots+w_n)}^{\frac{1}{2}}\,
\,\ktwo{\theta}{n+1} \langle \theta_{n}|.
\end{eqnarray} 
It is easily verified from these expressions that $[{\tilde a}_1, {\ad{\tilde a_1}}]$
is diag ($0,w_1,w_2,\cdots$) in the {\em transformed} basis, $i.e.,$
\begin{equation}
[{\tilde a}_1, {\ad{\tilde a_1}}] \,=\, \sum_{n=1}^{\infty} \,\,w_n
\,\,\ktwo{\theta}{n} \,\langle \theta_n |,
\end{equation}
as required ($cf$. Eq. (3.7)). 


In order to see how our general expressions simplify in
particular cases, it is
convenient to use Eq. (2.5) and the usual
representation of $a$ in the Fock basis, to express Eq. (3.14) in the form 
\begin{equation}
{\tilde a}_1 \;=\;\ad{b}\;\sum_{n=0}^{\infty}\;\left( \frac{W_{n+1}}
{(n+1)(n+2)}\right)^{\frac{1}{2}}
\,\kone{n} \langle {n+1}|\;b\;,
\end{equation}
or in the equivalent form
\begin{equation}
{\tilde a}_1 \;=\;\ad{b}\,(H+1)^{-\frac{1}{2}}\,\left \{ \sum_{n=0}^{\infty}\; W_{n+1}^{^{\frac{1}{2}}}
\,\kone{n} \langle {n+1}|\,\right \}\;(H+1)^{-\frac{1}{2}}\,b\;,
\end{equation}
where we have written 
\begin{equation}
W_n \,=\, w_1+w_2+\cdots +w_n
\end{equation}
for brevity.
It is clear from the above that, whenever
$W_{n+1}^{1/2}$ can be written as some function of
$n$, say $F(n)$, we obtain the closed-form expression 
\begin{equation}
{\tilde a}_1
\;=\;\ad{b}\,(H+1)^{-1}\,f(H)\,a\,(H+1)^{-\frac{1}{2}} \,b\;.
\end{equation}
If, further, $F(n)$ is a product of the form $(n+1)^{1/2}\,G(n)\,G(n+1)$, we
obtain the {\em symmetric} form 
\begin{equation}
{\tilde a}_1 \;=\;\ad{b}\,(H+1)^{-\frac{1}{2}}\,G(H)\,a\,G(H)\,(H+1)^{-\frac{1}{2}}\,b\;.
\end{equation}
Corresponding expressions obtain, of course, for $a_1^\dagger$ as well.
A remark is in order here on the square-root operator occurring in
these expressions. As $(\,1\,+\,H)^{-1}$ is a bounded positive definite
operator in $\cal H$, it has a unique positive square root in that
space, by virtue of the square-root lemma.$^{11}$\, Following the
procedure of Ref. 12, this can be defined
rigorously in terms of the resolvent operator
$(\,\xi\,+\,1\,+\,H)^{^{-1}}$ as
\begin{equation}
(\,1\,+\,H)^{- \frac{1}{2}}\; = \; \frac{1}{\pi}\;\int^{\infty}_{0}
\;d\xi\,{\;\xi^{-\frac{1}{2}}}\,(\,1\,+\,\xi\,+\,H)^{^{-1}}\;.
\end{equation}

We are now ready to read off  various  special cases of the
foregoing.

(i) First of all, if we set $w_n = w$ for {\em all} $n \ge 1$, the
operator in curly brackets in Eq. (3.18) above is simply $ w^{1/2} \,
a$, and we get
the symmetric expression
\begin{equation}
{\tilde a}_1 \;=\;w^{\frac{1}{2}}\;\ad{b}\,(H+1)^{-{\frac{1}{2}}}
\,a\,(H+1)^{-{\frac{1}{2}}}\,b.
\end{equation}

(ii) Next, we note that the `distorted' Heisenberg algebra
 introduced in Ref. 10 corresponds to setting
$w_1=w$ and $w_n = 1\;\;$ for $n \ge 2 $, so that
 $W_{n+1} = (w+n)$.
Therefore ${\tilde a}_1$ reduces, in this special case, to a closed-form 
expression -- namely, the
operator $C_w$  defined in Eqs. (2.14) and (2.15). 

(iii) Again, if we choose $w_n = n$ itself -- $i.e.,$ we have the algebra
\begin{equation}
[{\tilde a}_1, {\ad{\tilde a_1}}]\, =\, U\,H\,\ad{U} \,=\, \widetilde
H \,\,= \,\,\ad{b}b,
\end{equation}
 we get  
\begin{equation}
{\tilde a}_1 \;=\;2^{-\frac{1}{2}}\;\ad{b}\,(H+1)^{-\frac{1}{2}}
\,a\,b.
\end{equation}

(iv) On the other hand,
if we set $w_1=w$ and $w_n=0 \;\;$ for $n \ge 2 $, we get
\begin{equation}
{\tilde a}_1 \;=\;w^{\frac{1}{2}}\;\ad{b}\,(H+1)^{-1}
\,a\,(H+1)^{-{\frac{1}{2}}}\,b.
\end{equation}

(v) Finally, the $q$-generalization$\,^{13}$ of the isospectral
case is immediately found by setting $w_n = q^n$ : this implies that
\begin{equation}
[{\tilde a}_1, {\ad{\tilde a_1}}]\, =\, q^{\widetilde H},
\end{equation}
which is a transformed, equivalent, version of a $q$-commutation
relation between a conjugate pair of ladder operators. We obtain in this case 
the result
\begin{equation}
{\tilde a}_1 \;=\;q^{\frac{1}{2}}\;\ad{b}\,(H+1)^{-1}\,\left(
\frac{1-q^{H+1}}{1-q}\right )^{\frac{1}{2}}
\,\,a\,\,(H+1)^{-{\frac{1}{2}}}\,b.
\end{equation}
It is readily seen that Eq. (3.28) reduces to Eq. (3.23) (with $w$ =1) in
the limit $q \,\rightarrow \,1$.

\section*{\large IV. REPRESENTATION IN TERMS OF PSEUDO-DIFFERENTIAL OPERATORS}
\renewcommand{\theequation}{4.\arabic{equation}}
\setcounter{equation}{0}
\hskip .5cm The expressions obtained in the preceding section involve
various non-local operators, for which it is appropriate to exhibit
 suitable explicit representations.  Kernels in the position basis
for the integral operators represented by ${\tilde a}_1$ and
$\ad{{\tilde a}_1}$ may of course be written down at once from Eqs.
(3.14) and (3.15) in terms of the wavefunction ${\widetilde {\psi}}_n (x)
= \langle x | \theta_n
\rangle $ and its complex conjugate. 
{\em Formal} representations for the operators concerned in a form that
is an extension of the standard ones $a$ = $2^{^{-1/2}}\,(x+d\,)\,$, $\,\ad{a}$
= $2^{^{-1/2}}\,(x-d\,)$  may be given in terms of
pseudo-differential operators,$\,^{14,15}$\, $i.e.$, operators of the form
\begin{eqnarray}
\sum_{n=0}^{N} \, u_n(x) \,d^n \, + \,
\sum_{n=0}^{\infty} \, u_{-n}(x) \,d^{-n}\nonumber
\end{eqnarray}
where $N$ is a finite positive integer (and $d\equiv d/dx$). Here
$d^{-1}$ is the antiderivative, defined recursively through 
\begin{equation}
d^{-1}\,(f\,\cdot) \;=\; \;\sum^{\infty}_{n=0}\;(-1)^n \, f^{(n)}\, d^{-1-n}\,
(\cdot) 
\end{equation}
where $f^{(n)}$ is the $n$\,th derivative of $f(x)$. This leads to 
the basic relationship
\begin{equation}
[d^{-r}\,,\,f]\;=\;\sum_{n=1}^{\infty}\;(-1)^n\;
\left ( \begin{array}{c} n+r-1\\n-1 \end{array} \right ) f^{\,(n\,)} \,
\;d^{-n-r}\,.
\end{equation}
With the help of the foregoing we can find expansions for the
operators $(d^2\,+\,u(x))^{1/2}$ and $(d^2\,+\,u(x))^{-1/2}$ in
powers of $d^{-1}$. In particular, we find
\begin{eqnarray}
(1\,+\,H)^{-\frac{1}{2}}&=&-\,2^{\frac{1}{2}}\,\,i\,\,(d^2\,-\,x^2\,-1)
^{-\frac{1}{2}} \nonumber \\
&=& -\,2^{\frac{1}{2}}\,\,i\,\, \left \{\, d^{-1}\,+\,\frac{1}{2}\,(1\,+x^2)\,d^{-3}
\,-\,\frac{3}{2}\,x\,d^{-4}\,+\cdots\,\right \}.
\end{eqnarray}
Using these results, formal expansions for $\tilde a_1$ and $\ad{{\tilde a}_1}$
may be derived in those cases discussed in Sec. 3 in which closed-form 
expressions obtain for these operators. In particular,
corresponding to the algebra$\,^{10}$\, $[\tilde a_1 \,,\, \ad{{\tilde a}_1}]\;=\;
\mbox{diag}\,\, (w,1,1,\cdots)$ in $\widetilde {\cal H}_1$, for which
$\tilde a_1$  is the operator $C_w$ given by Eqs. (2.14) and (2.15), we
get the following expansions (quoted  up to $ O\,(\,d^{-2}$)\, for brevity)\,: 

\begin{equation}
2^{\frac{1}{2}}\,{\tilde
a}_1=x+d-(w-2-{\phi_{\lambda}}'\,)\,d^{-1}+\,[x\,(2-w)\,+\,\phi_{\lambda}\,{\phi_{\lambda}}'
\,+x\,{\phi_{\lambda}}' \, +\,2 \, \phi_{\lambda}\,]\,d^{-2}+\,\cdots
\end{equation}
and
\begin{equation}
2^{\frac{1}{2}}\,\ad{{\tilde a}_1}=x-d+(w\,-2\,-\,{\phi_{\lambda}}'\,)\,d^{-1}
+\,[\,x\,(2-w)\,-\,x\,{\phi_{\lambda}}'-\phi_{\lambda}\,{\phi_{\lambda}}'
\,]\,d^{-2}+\,\cdots\,.
\end{equation}
 Further, we find 
\begin{equation}
{\tilde a}_1\ad{{\tilde
a}_1}\;=\;\frac{1}{2}\,(-d^2\,+\,x^2\,+\,2w-3)\,-\,{\phi_{\lambda}}'\;,
\end{equation}
while in the space ${\mbox{span}}\,\{\,\ktwo{\theta}{n},\; n\ge 2\}$
we also have
\begin{equation}
\ad{{\tilde a}_1}{\tilde
a}_1\;=\;\frac{1}{2}\,(-d^2\,+\,x^2\,+2w-\,5)\,-\,{\phi_{\lambda}}'\;.
\end{equation}
When $w=1$, Eq. (4.8) is valid in all of $\widetilde {\cal H}_1$.

Finally, we remark that in the limit $|\lambda| \rightarrow \infty$,
$\phi_{\lambda}\,(x)$ and its derivatives vanish, $U \, \rightarrow
\,1$, $b \rightarrow a,\;\,
\widetilde {\cal H}_1\,\rightarrow  {\cal H}_1,\;\, {\tilde
a}_1\,\rightarrow a_1$, etc., in all the foregoing expressions.

\section*{\large V. COHERENT STATES}
\renewcommand{\theequation}{5.\arabic{equation}}
\setcounter{equation}{0}
\hskip .5cm With the operators $\tilde a_1$, $\ad{\tilde a_1}$ in
 ${\widetilde {\cal H}}_1$ and their commutation relation at hand
(Eqs. (3.14)-(3.16)), we may construct coherent states (CS) in the usual
manner, $i.e$., as eigenstates of $\tilde a_1$ built upon the base
state \ktwo{\theta}{1}. We find, for every $\zeta \in {\BC}$,
\begin{equation}
\tilde{a_1}  \kthree{\zeta}{\theta}{1} \;=\; \zeta \,\kthree{\zeta}{\theta}{1}
\end{equation}
where the normalized eigenstate is given by 
\begin{equation}
\kthree{\zeta}{\theta}{1} \,=\,h^{^{-\frac{1}{2}}}(|\zeta|^2
)\, \sum_{n=0}^{\infty} \,d_n^{^{\frac{1}{2}}} \,\zeta^n \, 
\ktwo{\theta}{n+1}
\end{equation}
with 
\begin{eqnarray}
d_0\;=\;1 \;,\; d_n &=& (W_1 \cdots W_n)^{-1}\,. 
\end{eqnarray}
The normalization factor in Eq. (5.2) is defined by 
\begin{equation}
h(|\zeta|^2) \,=\, \sum_{n=0}^{\infty}\; d_n \; |\zeta|^{2n}\;.
\end{equation}

A question of interest is the kind of Fock-Bargmann or
entire function representation of states$\,^{9,16}$ in ${\cal {\widetilde H}}_1$
associated with the CS constructed above -- specifically, the {\em
order}$\,^{17}$ of the class of entire functions involved. A state \kone{\Psi}
of finite norm is represented by the analytic function
\begin{equation}
\Psi(\zeta) \;=\; \sum_{n=0}^{\infty}\; d_n^{^{\frac{1}{2}}}\, \langle
\theta_{n+1}  \kone{\Psi}\; \zeta^n.
\end{equation}
As the state  \kthree{\zeta}{\theta}{1} in Eq. (5.2) is normalized to
unity, we have $|\Psi(\zeta)| \, \le \, ||\Psi|| \;|h|^{1/2}$.
Hence the  asymptotic behavior of $\Psi(\zeta)$ is decided by
that of $h^{^{\!1/2}}$ which, in turn, is
controlled by the growth properties of the sequence $\{ d_n\}$.
In particular, the {\em order} of the class of entire functions to
which $\Psi(\zeta)$ belongs is given by 
\begin{equation}
\rho \;=\; \lim_{n \rightarrow \infty} \; \frac{2n\,\log n}{\log \, (W_1 \cdots W_n)}.
\end{equation}
In the special cases (i)-(v) considered at the end of Sec. 3, this
leads to the following results: In Case (i), $(W_1 \cdots W_n)
\,=\,w^n\, n!$, while in Case (ii)$\,^{10}$ it is $\Gamma (n+w)/ \Gamma (n)$.
In both cases it follows easily from Eq. (5.6) that  $\rho\, =\,2$,
$i.e.$, the order is exactly the same as it is for the Fock-Bargmann
representation corresponding to the standard oscillator CS. In Case
(iii),  $(W_1 \cdots W_n) \,=\,2^{-n}\,n!\,(n+1)!$, leading to
$\rho\,=\,1$. All these cases are covered by the following
generalization: if the coefficient $w_n$ of the algebra $[\tilde
a_1\,,\,\ad{\tilde a_1}]\,=\,\mbox{diag}\,(w_1,w_2,\cdots)$ has the
leading asymptotic behavior $n^\nu$ for large $n$, with $\nu \,>
\,-1$, then $(W_1 \cdots W_n) \,\sim
\,n^{n(1+\nu)}$, which leads to the result 
\begin{equation}
\rho \,=\, 2 \,(1+\nu)^{-1}
\end{equation}
for the order of the corresponding entire function representation of
${\widetilde {\cal H}}_1$. Cases (i) and (ii) correspond to $\nu
\,=\,0$, while $\nu\,=\,1$ in Case (iii). In Case (iv),
$h\,=\, w\,(w\,-\,|\zeta|^2)^{-1}$, so that the
representation is not in terms of entire functions at all.

For the $q$-generalization, Case (v), $(W_1 \cdots W_n)$ is a
`$q$-factorial' given by 
\begin{equation}
q\,(q+q^2)\,\cdots\,(q+q^2+\cdots+q^n)\;=\; q^n\,(q-1)^{(1-n)}\,
(q^2-1)\,\cdots (q^n-1).
\end{equation}
When $q \,<\,1$, it is readily established that the series defining
$h$ has a finite radius of convergence ($|\zeta|\, =
\,q^{1/2} )$, as in Case (iv). When $q$ = 1, of course, we recover
Case (i) or (ii) with $w$ = 1, so that $\rho \,=\,2$. On the other
hand, when $q\,>\, 1$, $(W_1\cdots W_n)$ has the leading asymptotic
behaviour $q^{n^2/2}$, which implies that $\rho \,=\,0$.

The CS considered above have been constructed as annihilation
operator eigenstates. In the special case $w_n\,=\,1$ for all
$n\,\ge\,1$, they are also CS in the group-theoretic sense. Each
coefficient $c_n$ in Eq. (3.11) is now equal to unity, and the shift
operator $\widetilde S$ obeys the exact partial isometry condition
\begin{eqnarray}
\widetilde S \,\ad{\widetilde S} \;=\;1\;,\; \;\;\ad{\widetilde S}
\,\widetilde S \;=\; 1\,-\, \ktwo{\theta}{0} \,\langle \theta_0| \;.
\end{eqnarray}
We then have $[{\tilde a}_1, \ad{\tilde a_1}]\,=\,1$ in ${\widetilde
{\cal H}}_1$, and the Hamiltonian ${\widetilde  H}_1 \,=\, \ad{\tilde a_1}
{\tilde a}_1$ with eigenstates \ktwo{\theta}{n} and eigenvalues
$(n-1)$, $n\,=\,1,2,\cdots$, is the exact analog in ${\widetilde
{\cal H}}_1$ of the Hamiltonian  $H\,=\,\ad{a}a$ in $\cal H$. The
transformation from ($a$, $\ad{a}$) to $(\tilde a_1$, $\ad{\tilde
a_1}$) is a true operator map, in the sense that any function
$\phi\,(a\,,\,\ad{a}) \, \rightarrow \, \phi\,(\tilde a_1\,,\,\ad{\tilde
a_1})\,=\, \ad{\tilde S}\,\phi\,(a\,,\,\ad{a})\,\widetilde S$.
The coherent states 
\begin{equation}
\kthree{\zeta}{\theta}{1}\,\,=\,\,\exp\; 
\left(-{\frac{{|\zeta|}^2}{{\small 2}}}\,\right ) \,\,
\sum_{n=0}^{\infty}\,\,\frac{{\zeta}^n}{\sqrt{n!}}\,\,\ktwo{\theta}{n+1}
\,\,\,\,\,(\zeta\,\,\in\,\,{\BC})
\end{equation}
are   normalized eigenstates of the
annihilation operator ${\tilde a}_1$. They are {\em also}  CS in the Perelomov
sense, for we have \kthree{\zeta}{\theta}{1} = $D(\zeta,\overline{\zeta})
\ktwo{\theta}{1}$, where the displacement operator 
\begin{equation}
D(\zeta,\overline{\zeta}) = \exp\;(\zeta\, \ad{{\tilde a_1}} - \overline{\zeta} {\tilde a}_1)
\end{equation}
is unitary in ${\widetilde {\cal H}}_1$. Moreover, as in the case of the
standard oscillator CS, the state $\;$ \kthree{\zeta}{\theta}{1} is
the ground state of the displaced Hamiltonian
$ D \,{\widetilde H}_1\, \ad{D}\,$.    
The excited states \kthree{\zeta}{\theta}{n} $(n \geq 2)$ of
this Hamiltonian, $i.e.$, the generalized
coherent states corresponding to \kthree{\zeta}{\theta}{1} , are obtained
{\em either} by operating on \kthree{\zeta}{\theta}{1} by the
transformed raising operator ${(D \, {\ad{\tilde a_1}} \, \ad{D})}^{n-1}$, {\em
or} by displacing the corresponding base state \ktwo{\theta}{n}:
\begin{equation}
\kthree{\zeta}{\theta}{n} = {(D{\ad{\tilde a_1}} \ad{D})}^{n-1}
\kthree{\zeta}{\theta}{1} = D\,\ktwo{\theta}{n} \,\,, (n = 2,3, \ldots)\:.
\end{equation}
Thus, these coherent states and generalized CS have in ${\widetilde {\cal H}}_1$ the same
 standing as
their standard oscillator counterparts have in ${\cal H}$. They therefore constitute the closest
analogs of the latter for the isospectral family of oscillators
in ${\widetilde {\cal H}}_1$, while remaining a distinctly new (unitarily
inequivalent) class of coherent states. However, the generalized
algebra of Eq. (3.16) leads, as we have seen, to an even wider variety of possible
coherent states with interesting properties.

\vskip 2cm
\section*{ \large Acknowledgments}
\hskip .5cm We are grateful to Suresh Govindarajan for helpful discussions on pseudo-differential
operators.
SS acknowledges financial support from the Council of Scientific and Industrial Research,
India in the form of a Junior Research Fellowship.


\end{document}